\documentclass[aps,twocolumn,superscriptaddress,pra,longbibliography]{revtex4-2}
\usepackage[colorlinks=true, citecolor=blue, urlcolor=blue, linkcolor=red]{hyperref}
\usepackage{amssymb,booktabs,physics,orcidlink}

\begin{document}
\title{Strongly coupled giant-atom waveguide quantum electrodynamics}
\author{Zong-Wei Wu}
\affiliation{Key Laboratory of Quantum Theory and Applications of MoE, Lanzhou Center for Theoretical Physics, Gansu Provincial Research Center for Basic Disciplines of Quantum Physics, Key Laboratory of Theoretical Physics of Gansu Province, Lanzhou University, Lanzhou 730000, China}
\author{Jun-Hong An\orcidlink{0000-0002-3475-0729}}\email{anjhong@lzu.edu.cn}
\affiliation{Key Laboratory of Quantum Theory and Applications of MoE, Lanzhou Center for Theoretical Physics, Gansu Provincial Research Center for Basic Disciplines of Quantum Physics, Key Laboratory of Theoretical Physics of Gansu Province, Lanzhou University, Lanzhou 730000, China}

\begin{abstract}
Describing systems of superconducting atoms coupled to a continuum of photonic modes at multiple separated locations in a waveguide, waveguide quantum electrodynamics (QED) with giant atoms has emerged as a promising platform for realizing quantum interconnect. Such systems have been reported to exhibit rich phenomena that differ from those of natural atoms. Going beyond the widely used Born-Markov and Wigner-Weisskopf approximations, we investigate the non-Markovian dynamics of one and two giant atoms interacting with a waveguide formed by an array of coupled resonators. We discover that the diverse dynamical behaviors of the giant atoms are intrinsically determined by the energy spectrum of the composite system consisting of the giant atoms and the photonic modes in the waveguide. As long as one and more bound states are present in the energy spectrum, their excited-state probabilities, respectively, tend to stable finite values and lossless Rabi-like oscillations with frequencies proportional to the differences of the bound-state eigenenergies. Our result provides an insightful guideline for suppressing the decoherence of giant atoms and facilitates the development of quantum interconnect devices using giant-atom waveguide QED.
\end{abstract}
\maketitle

\section{Introduction}

The field of waveguide quantum electrodynamics ($\mathrm{QED}$) has been significantly advanced by the introduction of artificial giant atoms typically realized with superconducting circuits in the microwave regime \cite{You2011,PhysRevA.76.042319,10.1007/978-981-15-5191-8_12,Wallraff2004,PhysRevA.103.023710,GU20171,PhysRevLett.121.090502,PhysRevResearch.7.L012036,Zanner2022}. Their excellent compatibility with hybrid quantum systems \cite{Aref2016,Manenti2017,Andersson2019,doi:10.1126/sciadv.1603150,Clerk2020,Wang2022,Wang2022} underscores considerable potential for both fundamental research \cite{Storz2023,PhysRevLett.126.141302} and the development of quantum technologies such as quantum computing \cite{Blais2020,PhysRevX.11.021010,Arute2019}, sensing \cite{doi:10.1126/science.aaz9236,Johnson2010}, and routing \cite{PhysRevLett.107.073601}. In contrast with natural atoms modeled with a point-like dipole, giant atoms coupled to a waveguide at multiple spatially separated points lead to a breakdown of the dipole approximation and give rise to a rich spectrum of quantum effects \cite{Manenti2017,PhysRevLett.119.180505,PhysRevLett.120.223603,PhysRevA.101.053855}. These effects include frequency-dependent Lamb shifts \cite{PhysRevA.90.013837} and relaxation rates \cite{PhysRevA.103.023710,PhysRevLett.126.043602,PhysRevLett.128.223602}, decoherence-free interactions \cite{Kannan2020,PhysRevLett.120.140404,PhysRevResearch.2.043184,PhysRevA.105.023712}, nonexponential decay \cite{PhysRevA.95.053821,Andersson2019,PhysRevA.102.033706}, and the formation of unconventional bound states \cite{PhysRevLett.126.043602,PhysRevResearch.2.043014,PhysRevA.102.033706,PhysRevA.104.053522,PhysRevA.107.023716,PhysRevA.108.013704}. These remarkable properties have motivated extensive theoretical investigations \cite{Longhi:20,Xiao2022,PhysRevA.106.033522,PhysRevA.107.013710,PhysRevResearch.6.013279}, which have charted a clear path toward utilizing giant-atom platforms to achieve chiral quantum optics \cite{PhysRevA.107.063703,Chen2022,PhysRevResearch.3.043226,PhysRevLett.127.233601,Wang_2022,ZHANG2023170568} and simulate many-body physics \cite{PhysRevA.111.023712,PhysRevResearch.3.033233}.

Waveguide QED with giant atoms has emerged as a highly promising platform for realizing quantum interconnects of different nodes \cite{PRXQuantum.2.017002}. Quantum interference among multiple coupling pathways of giant atoms with the photonic modes in the waveguide enables an unprecedented capacity in controlling light–matter interactions, offering an efficient control dimension for generating and stabilizing quantum entanglement between separated giant atoms \cite{PhysRevLett.130.053601,PhysRevA.108.023728,PhysRevA.104.013720,PhysRevLett.120.140404}. A widely used method to treat the interactions between the giant atoms and the photonic modes in the waveguide is the Born-Markov approximation using the SLH (scattering operator, Lindblad operators, Hamiltonian) formalism for cascaded quantum systems \cite{PhysRevLett.120.140404,PhysRevLett.130.053601,Combes04052017,Kannan2020,PhysRevResearch.5.043135,PhysRevX.13.021039,PhysRevA.108.023728,PhysRevA.104.013720,PhysRevA.105.023712,PhysRevLett.133.063603}. However, when the photon propagation time between different coupling points becomes comparable to the typical timescale of the giant atoms, the system exhibits the non-Markovian effect. Theoretical frameworks such as the Wigner–Weisskopf (WW) approximation \cite{PhysRevA.95.053821,PhysRevResearch.2.043014,PhysRevA.108.013704,PhysRevA.106.063703,PhysRevA.106.063717,PhysRevResearch.4.023198,Xu_2024} have been employed to describe such systems, capturing the interplay between quantum interference and delayed feedback. This method partially reflects the non-Markovian effect on the dynamics. However, it generally works only under the condition that the couplings between the giant atoms and the photonic modes are weak. The rapid development of waveguide QED with giant atoms, particularly in the strong-coupling regime \cite{Liu2017}, necessitates an exactly non-Markovian framework that can consistently reveal the rich dynamical features and their intrinsically dominant rule.

In this paper, we move beyond the conventional approximations to explore the non-Markovian dynamics of one and two giant atoms interacting with a waveguide formed by a coupled resonator array. Our study reveals a one-to-one correspondence between the diverse features of the energy spectrum of the composite giant-atom-photon system and the rich dynamical evolution behaviors of the giant atoms. We find that the excited-state population of the giant atoms either saturates at a finite value or exhibits a persistently lossless Rabi oscillation. It is essentially governed by different numbers of bound states, including both in and out of the continuum, formed in the energy spectrum of the atom-photon system. Especially, this mechanism, underpinned by bound-state formation offers a powerful approach for suppressing decoherence and generating steady-state entanglement between distant giant atoms. Our findings provide profound insights into decoherence mitigation and pave the way for designing more extensive quantum devices based on giant-atom waveguide QED.

The rest of this paper is organized as follows: In Sec. \ref{sec_II}, we present the Hamiltonian of the system consisting of $N$ giant atoms coupled to a structured waveguide and develop a general formalism for its non-Markovian dynamics and eigenenergy spectrum. In Sec. \ref{sec_III}, we employ this theoretical framework to analyze the single giant-atom system, elucidating the connection between its energy-spectrum feature and the dynamical behaviors. These findings are then generalized to the case of two giant atoms in Sec. \ref{sec_IV} to demonstrate the efficacy of this mechanism in generating stable entanglement between the distant giant atoms. We discuss the experimental feasibility and summarize the main conclusion in Sec. \ref{sec_V}.

\section{Model and dynamics}\label{sec_II}
\begin{figure}[tbp]
\includegraphics[width=1\columnwidth]{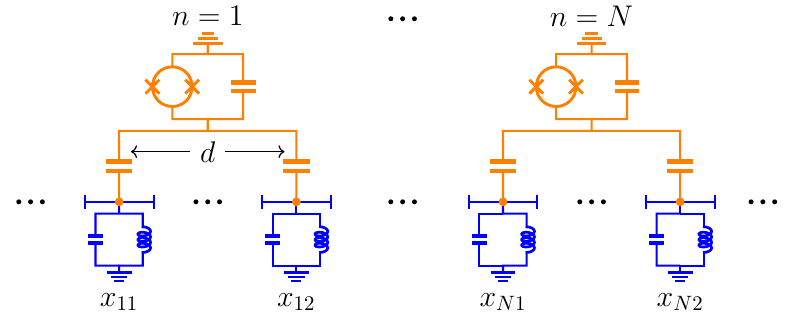}
\caption{Schematic diagram of $N$ giant atoms in orange interacting with $M=2$ resonators in blue of a one-dimension nearest-neighbor coupled-resonator array realized in circuit QED system. $d$ is the distance between the two resonators coupling to the giant atoms. $x_{nm}$ is the position of the $m$th resonator interacting with the $n$th giant atom. }\label{fig1}
\end{figure}
	
We consider that $N$ giant atoms formed by artificial superconductor atoms with identical frequency $\Omega$ and described by the Hamiltonian $\hat{H}_\text{S}=\Omega\sum_{n=1}^N\hat{\sigma}_n^\dag\hat{\sigma}_n$ interact with the electromagnetic environment in a one-dimensional waveguide. The waveguide is composed of an array of coupled LC-circuit resonators. Its Hamiltonian is $\hat{H}_\text{E}=\omega_c\sum_{l=1}^L\hat{\tilde a}_l^\dag\hat{\tilde a}_l-h\sum_{l=1}^{L-1}(\hat{\tilde a}_l^\dag\hat{\tilde a}_{l+1}+\text{H.c.})$, where $\omega_c$ is the frequency of $L$ identical resonators, $\hat{\tilde a}_l$ is the annihilation operator of the $l$th resonator, and $h$ is their nearest-neighbor coupling strength. Performing the Fourier transform, we obtain $\hat{H}_\text{E}=\sum_k(\omega_c-2h\cos k)\hat{a}_k^\dag\hat{a}_k$, where $\hat{a}_k$ is the Fourier transform of $\hat{\tilde a}_l$ and $k=2n\pi/L$ for $n=1,2,..., L$. In contrast with a natural atom that interacts with only a single resonator of the waveguide, each giant atom is able to interact with the environment through $M$ different resonators, see Fig. \ref{fig1}. In the rotating frame of \begin{equation}\hat{H}_0=\omega_c[\sum_{n=1}^N\hat{\sigma}_n^{\dagger} \hat{\sigma}_n+\sum_k\hat{a}^\dag_k\hat{a}_k],
\end{equation} the Hamiltonian of the total system becomes \cite{PhysRevLett.126.043602,PhysRevA.108.013704}
\begin{eqnarray}
\hat{H}=\Delta\sum_{n=1}^N \hat{\sigma}_n^{\dagger} \hat{\sigma}_n+\sum_k[ \omega_{k}\hat{a}_k^\dagger \hat{a}_k+\sum_{n=1}^N(g_{kn}\hat{a}_{k}\hat{\sigma}_{n}^{\dagger}+\rm{H.c.} )],\label{eq_H}~~~~
\end{eqnarray}
where $\Delta=\Omega-\omega_c$, $\omega_k=-2h\cos k$, $g_{kn}=g_0\sum_{m=1}^Me^{ikx_{nm}}/\sqrt{L}$, and $x_{nm}$ is the position of the $m$th resonator interacting with the $n$th giant atom. An indirect coupling between different giant atoms is mediated by their direct coupling to the common photonic modes in the waveguide. This is characterized by the so-called spectral density $J_{nn'}(\omega)=\sum_kg_{kn}g_{kn'}^*\delta(\omega-\omega_k)$. It can be analytically proven that $J_{nn'}(\omega)=J_{n'n}(\omega)\equiv J_{|n-n'|}(\omega)$.

First, the system has a rich energy-spectrum structure. The total excitation number of the system $\hat{\mathcal N}=\sum_n\hat{\sigma}_n^\dag\hat{\sigma}_n+\sum_k\hat{a}^\dag_k\hat{a}_k$ is conserved due to $[\hat{H},\hat{\mathcal N}]=0$. The eigenstate of $\hat{H}$ in the single-excitation subspace is expanded as
\begin{equation}|\Phi\rangle=[\sum_n \alpha_n\hat{\sigma}_n^\dag+\sum_k\beta_k\hat{a}_k]|g^{\otimes N},\{0_k\}\rangle,
\end{equation}
where $|g\rangle$ is the ground state of the giant atom and $|\{0_k\}\rangle$ is the vacuum state of the environment. From $\hat{H}|\Phi\rangle=E|\Phi\rangle$, we derive
\begin{eqnarray}
(E-\Delta)\alpha_n=\sum_k g_{kn}\beta_k,\label{altt}\\
(E-\omega_k)\beta_k=\sum_{n'} g_{kn'}^*\alpha_{n'}\label{btk}.
\end{eqnarray}
Substituting the solution of Eq. \eqref{btk} as $\beta_k=\sum_{n'}g_{kn'}^*\alpha_{n'}/(E-\omega_k)$ into Eq. \eqref{altt}, we obtain a matrix equation $[E-\Delta-\int_{-2h}^{2h} d\omega{{\bf J}(\omega)\over E-\omega}]{\pmb \alpha}=0$ in the continuous limit of the environmental frequency, where ${\pmb \alpha}=(\alpha_1,...,\alpha_N)^\text{T}$ and $\mathbf{J}(\omega)$ is a $N$-by-$N$ matrix formed by the elements $J_{nn'}(\omega)$. The equation has nontrivial solutions $E$ if and only if
\begin{equation}
\text{det}[E-\Delta-\int_{-2h}^{2h} d\omega{{\bf J}(\omega)\over E-\omega}]=0,\label{eigene}
\end{equation}
from which the energy spectrum of the total system consisting of the giant atoms and the environment is determined. The energy-spectrum equation \eqref{eigene} generally has three types of solutions \cite{Yang2025}. The first one is an infinite number of solutions falling in the environmental continuous-band regime $(-2h,+2h)$, which makes the integral in Eq. \eqref{eigene} ill-defined due to the singularity of its integral function. They form a continuous energy band in the energy spectrum of the total system. The second one is possibly formed isolated solutions $E^\text{boc}_j$ in the regimes $(-\infty,-2h)$ and $(2h,+\infty)$. Since they fall out of the continuous band, we call the corresponding eigenstates bound states out of the continuum (BOCs). The third one is several discrete solutions $E^\text{bic}_j$ in the regime $(-2h,+2h)$ that are the removable singularities of the integral function in Eq. \eqref{eigene}. Since they fall in the continuous energy band, we call the corresponding eigenstates bound states in the continuum (BICs).

Second, we study the dynamics of the giant atoms under the initial condition that only the first giant atom is in the excited state and the environment is in the vacuum state. The evolved state of the total system is expanded as $|\Psi(t)\rangle=[\sum_{n} c_{n}(t)\hat{\sigma}^\dag_n+\sum_kd_k(t)\hat{a}^{\dagger}_k]|g^{\otimes N},\{0_k\}\rangle$. From the time-dependent Schr\"{o}dinger equation, we obtain the equation satisfied by the column vector ${\bf c}(t)=(c_1(t),...,c_N(t))^\text{T}$ as \cite{PhysRevLett.132.090401}
\begin{equation}
\dot{\bf c}(t)+i\Delta{\bf c}(t)+\int_0^t{\bf G}(t-\tau){\bf c}(\tau)d\tau=0,\label{intfgr}
\end{equation}
where ${\bf G}(t-\tau)=\int_{-2h}^{2h}d\omega{\bf J}(\omega)e^{-i\omega (t-\tau)}$ is a $N$-by-$N$ matrix of the environmental correlation functions. Manifesting the memory effect, the convolution in Eq. \eqref{intfgr} renders the dynamics non-Markovian. Although Eq. \eqref{intfgr} is only solvable by numerical calculation, its long-time form can be analytically derived by the Laplace transform $\tilde{F}(s)=\int^{\infty}_0F(t)e^{-st}dt$. It converts Eq. \eqref{intfgr} into $\tilde{\mathbf{c}}(s)= [s+i\Delta+\tilde{\mathbf{G}}(s)]^{-1}\textbf{c}(0)$, where $\tilde{\bf G}(s)=\int_{-2h}^{2h}d\omega{{\bf J}(\omega)\over s+i\omega}$ is the Laplace transform of ${\bf G}(t-\tau)$. ${\bf c}(t)$ is derived by making the inverse Laplace transform to $\tilde{\bf c}(s)$, which requires finding the poles of $\tilde{\mathbf{c}}(s)$. It is remarkable to find that the pole equation of $\tilde{\mathbf{c}}(s)$ is just Eq. \eqref{eigene} after replacing $s$ with $-iE$. It indicates that the dynamics of the giant atoms characterized by ${\bf c}(t)$ is intrinsically determined by the features of the energy spectrum of the total system consisting of the giant atoms and their environment. With the poles in hands, the inverse Laplace transform is evaluated using the Cauchy’s residue theorem and contour integration as \cite{PhysRevLett.109.170402}
\begin{equation}
\mathbf{c}(t)=\mathbf{Z}(t)+\int_{0}^{\infty}\frac{d\omega }{2\pi }[\mathbf{\tilde{c}}(0^{+}-i\omega)-\mathbf{\tilde{c}}(0^{-}-i\omega )]e^{-i\omega t},\label{sol}
\end{equation}
where $\mathbf{Z}(t)=\sum_{\alpha=\text{bic,boc}}\sum_{j=1}^N$Res$[\mathbf{\tilde{c}}(-iE _{j}^{\alpha})]e^{-iE _{j}^{\alpha}t}$ and $\text{Res}[\mathbf{\tilde{c}}(-iE _{j}^{\alpha})]$ is the residue contributed by the $j$th bound state. The second term comes from the continuous energy band. Containing an infinite number of superposition components oscillating with time in continuously changing frequencies $\omega$, it tends to zero in the long-time limit due to the out-of-phase interference. Thus, if the bound state is absent, then $\lim_{t\rightarrow\infty}\mathbf{c}(t)={\bf 0}$ characterizes a complete decoherence, while if the bound states are formed, then $\lim_{t\rightarrow\infty}\mathbf{c}(t)=\mathbf{Z}(t)$ implies a suppression of decoherence.

The dominant role played by the energy-spectrum feature in the reduced dynamics of the giant atoms would be destroyed by both the widely used Born-Markov and WW approximations \cite{PhysRevA.81.052330}. Under the physical assumptions that the atom-environment coupling is weak and the correlation time of the environment is significantly shorter than the characteristic timescale of the atoms, we make the WW approximation by replacing ${\bf c}(\tau)$ in Eq. \eqref{intfgr} by ${\bf c}(t)$ \cite{PhysRevResearch.2.043014,PhysRevA.108.013704}. Neglecting the memory effect incorporated in the convolution, it is also called the first Markov approximation. Then the solution is
\begin{equation}
{\mathbf c}_\text{WW}(t)=e^{-i\Delta t-\int_0^td\tau \int_0^\tau d\tau' \mathbf{G}(\tau')}\mathbf{c}(0).\label{cww}
\end{equation}
Further extending the upper limit of the $\tau$-integral in Eq. \eqref{intfgr} from $t$ to infinity, which is called the second Markov approximation, we obtain the Born-Markov approximate solution
\begin{eqnarray}
\mathbf{c}_\text{MA}(t)=e^{-[\pi{\bf J}(\Delta)+i\bar{\pmb \Delta} ]t}\mathbf{c}(0),
\end{eqnarray}where $\bar{\pmb\Delta}=\Delta+\mathcal{P}\int_{-2h}^{2h} d\omega {{\bf J}(\omega)\over \Delta-\omega}$ and $\mathcal{P}$ denotes the Cauchy principal value. It indicates that the excited-state population $|c_\text{MA}(t)|^2$ of the giant atom experiences an exponential decay to zero due to the positive semidefiniteness of the matrix ${\bf J}(\Delta)$ under the condition $\Delta\in(-2h,2h)$. Note that some values of $\Delta$ may make ${\bf J}(\Delta)$ possess a vanishing eigenvalue, where the decay of $|{\bf c}_\text{MA}(t)|^2$ stops even under the Born-Markov approximation.
	
\section{Single giant atom}\label{sec_III}
We first investigate a giant atom that interacts with the waveguide through two resonators, i.e., $N=1$ and $M=2$. Then the spectral-density matrix only has one component
\begin{equation}
J_{0}(\omega)={g_0^2\over \pi} {1+T_d(\tfrac{-\omega}{2h})\over \sqrt{4h^2-\omega^2}},\label{spcrd}
\end{equation}
where $|x_{n2}-x_{n1}|=d>0$ is the separation between the two resonators coupled to the giant atom and $T_{n}(x)=\cos(n\arccos x)$ is the $n$th Chebyshev polynomial. The spectral density $J_0(\omega)$ becomes divergent at the edges of the band $\pm 2h$, which is called van Hove singularities \cite{PhysRev.89.1189}. The eigen equation \eqref{eigene} becomes
\begin{equation}
   Y(E)=E,~~~~Y(E)=\Delta+\int_{-2h}^{2h}d\omega {J_0(\omega)\over E-\omega}.\label{eisptm}
\end{equation}
First, it is easy to find that the eigen equation has an infinite number of roots in $|E|\le 2h$, which form a continuous energy band. Second, $Y(E)$ is a decreasing function when $|E|>2h$. Thus, a BOC with eigenenergy $E^\text{boc}>2h$ or $E^\text{boc}<-2h$ is present as long as $ Y(2h)>2h$ or $Y(-2h)<-2h$. It is proven analytically that
\begin{eqnarray}
\lim_{E\rightarrow 2h}Y(E)&=&\Delta+\left\{
		\begin{aligned}
			&+\infty,~d\in \mathbb{E}\\
			&dg_0^2/h,~d\in \mathbb{O},\\
		\end{aligned}
		\right.\label{boc11}\\
    \lim_{E\rightarrow -2h}Y(E)&=&-\infty. \label{boc12}
\end{eqnarray} Thus, we always have a BOC with $E^\text{boc}>2h$ for an even $d$ and a BOC with $E^\text{boc}<-2h$ for any system parameters. Being caused by the divergence of the $J_0(\omega)$ at the two van Hove singularities $\pm2h$, their eigenenergies asymptotically merge to the continuous energy band. We call them type-I BOCs. For an odd $d$, another BOC with $E^\text{boc}>2h$ suddenly appears when
\begin{equation}
  \Delta+dg_0^2/h>2h,  \label{bsII}
\end{equation} which removes the influence of the van Hove singularity at $2h$ on the energy spectrum. We call it type-II BOC. Third, there are discrete removable singularities $E_l^\text{bic}$ in the regime $(-2h,2h)$ determined by $J_0(E^\text{bic})=0$, which results in
\begin{equation}
E_l^\text{bic}=-2h\cos[(2l+1)\pi/ d], \label{bicccd}
\end{equation}
with $l$ being integer numbers. It can be analytically proven that $Y(E_l^\text{bic})=\Delta$. Thus, the eigen equation becomes $\Delta=E^\text{bic}_l$, which, combined with Eq. \eqref{bicccd} gives the condition of the system parameters for forming the BICs. The requirement of the presence of the BICs $J_0(E^{\text{bic}})=0$ is fundamentally caused by the destructive interference between the two interaction pathways of the giant atom and the common electromagnetic field in the waveguide.

It is easy to derive that the Laplace transform of Eq. \eqref{intfgr} for $N=1$ reads
\begin{equation}
\tilde{c}(s)=[s+i\Delta +\int_{-2h}^{2h}d\omega {J_0(\omega)\over s+i\omega}]^{-1}.
\end{equation} Substituting the eigenenergies of the bound states into Eq. \eqref{sol}, the steady-state solution of $c(t)$ in the non-Markovian dynamics is obtained as
\begin{equation}
    c(\infty)=\sum_{\alpha=\text{bic,boc}}\sum_{j}Z^\alpha_je^{-iE_j^{\alpha}t},\label{stdfm}
\end{equation}where $Z^\alpha_j=[\partial_s\tilde{c}(s)^{-1}|_{s\rightarrow-iE_j^\alpha}]^{-1}$. This is the exact long-time solution of $c(t)$ without any approximation from Eq. \eqref{intfgr}. Note that, if the eigenenergy $E^\text{boc}$ of type-I BOC is very close to the energy band, then its contributed residue is small and its dynamical effect is weak. This is in sharp contrast with the type-II BOC.

\begin{figure}[tbp]
\includegraphics[width=1.03\columnwidth]{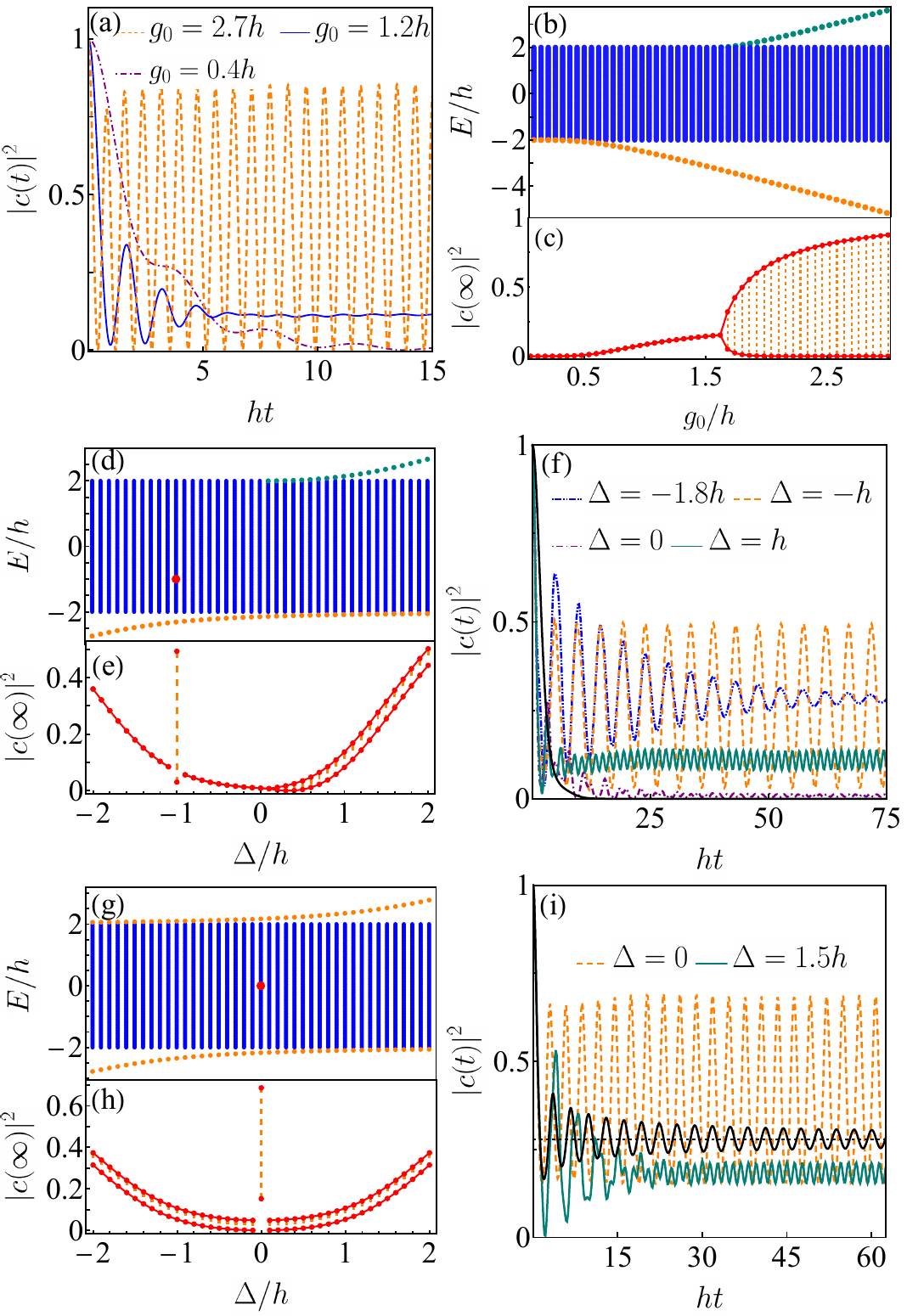}
\caption{(a) Non-Markovian evolution of the excited-state population $|c(t)|^2$, (b) energy spectrum, and (c) steady-state solution $|c(\infty)|^2$ in different $g_0$ when $\Delta=-0.6h$ and $d=1$. (d) Energy spectrum, (e) steady-state solution $|c(\infty)|^2$, and (f) non-Markovian evolution of $|c(t)|^2$ in different $\Delta$ when $g_0=0.8h$ and $d=3$. (g) Energy spectrum, (h) steady-state solution $|c(\infty)|^2$, and (i) non-Markovian evolution of $|c(t)|^2$ in different $\Delta$ when $g_0=0.8h$ and $d=2$. In planes (c), (e) and (h), the dots are obtained from numerically solving Eq. \eqref{intfgr}, the solid lines are from the analytical solution in Eq. \eqref{stdfm}, and the orange dots cover the values during its persistent oscillation. The red dot in planes (d) and (g) denotes the energy of BIC. The black solid lines in (f) and (i) are the WW approximate results from Eq. \eqref{wwapr}. We use $L=200$.}\label{fig2}
\end{figure}

The WW approximate solution is straightforwardly derived from Eq. \eqref{cww} as
\begin{eqnarray}
c_\text{WW}(\infty)=\left\{
		\begin{aligned}
			&\exp [-i\Delta t-{(2l+1)g_0^2\over h^2}],~d=4l+2\\		  &0,\text{others}.\\
		\end{aligned}
		\right. ~~~~\label{wwapr}
\end{eqnarray}
Comparing Eq. \eqref{wwapr} with Eq. \eqref{stdfm}, we see that the dominant role played by the bound states in the exact dynamics is missed by the WW approximate solution.

The dominant role played by the bound states of the total system in the dynamics of the giant atoms is further elaborated as follows. The dynamically evolved state $|\Psi(t)\rangle$ can alternatively be solved by the spectral decomposition method. Expanding it in the complete basis formed by the eigenstates of the total system, we have
\begin{equation}
    |\Psi(t)\rangle=\sum_{\alpha=\text{bic,boc}}\sum_jx_j^\alpha e^{-iE_j^\alpha t}|\Phi_j^\alpha\rangle+\sum_{j=\text{band}}y_j e^{-iE_j t}|\Phi_j\rangle,\label{bandexp}
\end{equation}where $x_j^\alpha=\langle\Phi_j^\alpha|\Psi(0)\rangle$ and $y_j=\langle\Phi_j|\Psi(0)\rangle$. Then the excited-state probability amplitude $c(t)$ of the giant atom is obtained as $c(t)=\langle e,\{0_k\}|\Psi(t)\rangle$. The contribution of the last term of Eq. \eqref{bandexp} tends to zero with time due to the out-of-phase interference under the continuously changing oscillation frequency $E_j\in[-2h,2h]$. Therefore, only the bound-state components survive in the long-time limit. This is the physical mechanism through which the dynamical behaviors of the giant atom are governed by the energy-spectrum features of the total system. However, this governess is achievable only under the exact non-Markovian dynamics. As long as any approximation, either Born-Markov or WW approximation, is made in the dynamics, it would be lost.

We plot in Fig. \ref{fig2}(a) the time evolution of the excited-state population $|c(t)|^2$ of the giant atom in different values of the coupling strength $g_0$ when $d=1$ and $\Delta=-0.6h$ by numerically calculating Eq. \eqref{intfgr}. The exact dynamics exhibits rich behavior, including the asymptotical damping to zero when $g_0=0.4h$, the stabilization to a constant when $g_0=1.2h$, and the periodic oscillation when $g_0=2.7h$. Solving Eq. \eqref{eisptm} numerically, we obtained the energy spectrum of the total system [see Fig. \ref{fig2}(b)]. It indicates that two branches of BOC energies separate the spectrum into three regions, i.e., the regions without BOC when $g_0\le 0.5h$, one BOC when $g_0\in (0.5,1.6]h$, and two BOCs when $g_0>1.6h$. The lower branch belongs to the type-I BOC, which is caused by the van Hove singularity of $Y(E)$ at $E=-2h$. The upper branch belongs to the type-II BOC, whose formation is consistent with the condition \eqref{bsII}. Figure \ref{fig2}(c) reveals that the steady-state population numerically calculated from Eq. \eqref{intfgr} shows good agreement with the one evaluated from the energies of the BOCs in Eq. \eqref{stdfm}. This proves the one-to-one correspondence between the rich dynamical behaviors and the energy-spectrum features. The dominant role played by the bound states in the dynamics is confirmed by the results in different $\Delta$ when $g_0=0.8h$ and $d=3$, see Fig. \ref{fig2}(d). The type-I BOC with $E^\text{boc}<-2h$ always exists. The type-II BOC is abruptly formed when $\Delta>0.08h$. An exception occurs at $\Delta=-h$, where a BIC obeying the condition \eqref{bicccd} is present. $|c(t)|^2$ evolves to a nonzero value in the regime with one BOC and to a lossless oscillation with frequency $|E^\text{boc}_1-E^\text{boc}_2|$ or $|E^\text{boc}-E^\text{bic}|$ in the regimes with two BOCs or one BIC and one BOC. The numerical results match those evaluated from $E^\text{boc}$ and $E^\text{bic}$ via Eq. \eqref{stdfm}, see Fig. \ref{fig2}(e). However, these rich dynamical behaviors are not captured by the WW approximation \cite{PhysRevResearch.2.043014,PhysRevA.108.013704}, under which the results exclusively show the damping to zero irrespective of the value of $\Delta$ [see the black solid line in Fig. \ref{fig2}(f)].

For $d=2$, Fig. \ref{fig2}(g) confirms that the two type-I BOCs are always present for any $\Delta$. The excited-state population evaluated from Eq. \eqref{intfgr} in the long-time limit matches well the analytical result in Eq. \eqref{stdfm} [see Fig. \ref{fig2}(h)]. The transient evolution of $|c(t)|^2$ exhibits a lossless oscillation with frequency $|E^\text{boc}_1-E^\text{boc}_2|$ in the whole regime of $\Delta$ [see Fig. \ref{fig2}(i)]. An exception occurs at $\Delta=0$, where a BIC obeying the analytical condition \eqref{bicccd} is present. In this case, we have two BOCs with energies $E^\text{boc}_1=-E^\text{boc}_2\equiv \mathcal{E}$ and one BIC with energy $E^\text{bic}=0$. It is thus expected from Eq. \eqref{stdfm} that the three bound states contribute to $|c(\infty)|^2$ the lossless oscillation with frequencies $|E_{1/2}^\text{boc}-E^\text{bic}|=\mathcal{E}$ and $|E_{1}^\text{boc}-E_{2}^\text{boc}|=2\mathcal{E}$. Thus the trifrequency oscillation in this case reduces to a single-frequency one, as verified by the numerical result in Fig. \ref{fig2}(i). Such rich dynamical behaviors in different values of $\Delta$ are missed by the WW approximation, which asymptotically tends to a constant matching Eq. \eqref{wwapr} irrespective of the value of $\Delta$ [see the black solid line in Fig. \ref{fig2}(i)].

\begin{figure}[tbp]
\includegraphics[width=\columnwidth]{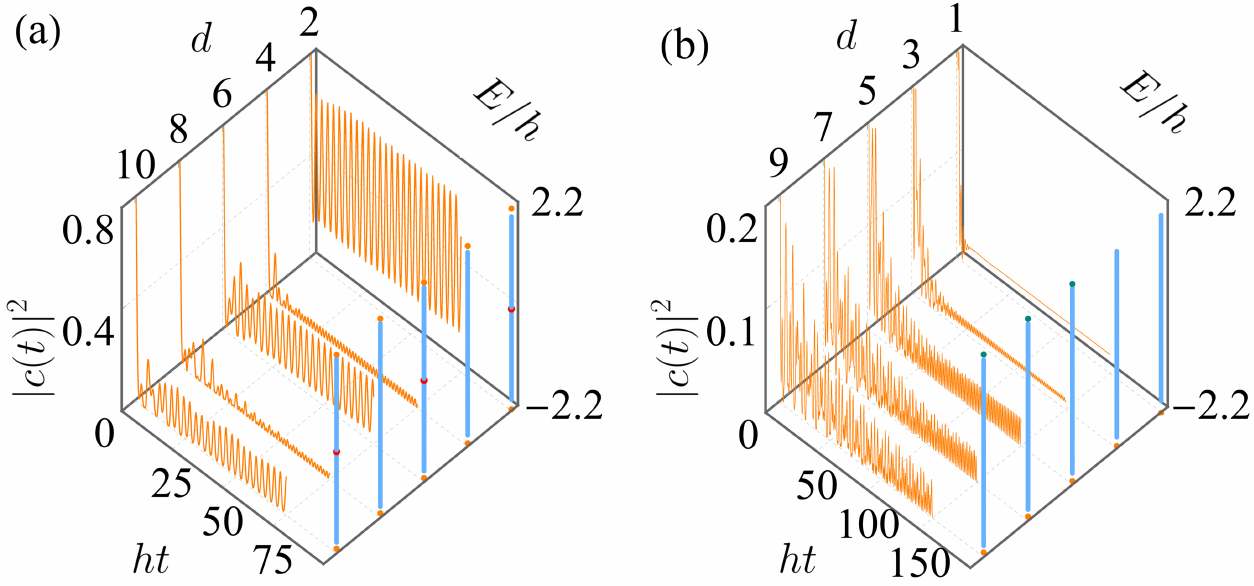}
\caption{Non-Markovian evolution of $|c(t)|^2$ and energy spectrum for (a) even and (b) odd $d$. The BIC (red dots) is present when $d=2$, $6$, and $10$. For even $d$, the type-I BOC (orange dots) is present in the above and the below of the band. For odd $d$, the type-I BOC (orange dots) is only present in the below of the band and the type-II BOC (light-blue dots) is present when $d\ge 5$. We use $\Delta=0$ and the other parameter values are the same as in Fig. \ref{fig2}(g). }\label{fig3}
\end{figure}

Figure \ref{fig3} shows the energy spectrum and the evolution of $|c(t)|^2$ in different values of $d$ when $\Delta=0$ and $g_0=0.8h$. When $d=2$, $6$, and $10$, there is always a proper integer number $l$ such that the condition \eqref{bicccd} for the existence of BIC is satisfied, as confirmed by Fig. \ref{fig3}. According to Eqs. \eqref{boc11} and \eqref{boc12}, the type-I BOC is present both above and below the continuous energy band for even $d$, while it is present only below the band for odd $d$. According to Eq. \eqref{bsII}, the type-II BOC is present only for the odd $d\ge 5$. These analytical conclusions are confirmed by the numerical results in Fig. \ref{fig3}. The type-I BOC is caused by the van Hove singularity of $J_0(\omega)$ at $\pm2h$. In contrast, when the two interaction pathways of the giant atom and the common electromagnetic field in the waveguide interfere constructively for odd $d$, the divergence of $Y(E)$ caused by the van Hove at $E=2h$ of $J_0(\omega)$ vanishes and $Y(2h)$ reaches a finite value. According to the condition of the formation of the BOC, as long as $Y(2h)>2h$, a type-II BOC appears. Being absent in the small-atom system, the type-II BOC is unique for the giant-atom system. The exact dynamics evaluated by numerically solving Eq. \eqref{intfgr} verifies that $|c(t)|^2$ tends to a constant when one BOC is formed, a lossless oscillation in a frequency $|E_1^\text{boc}-E_2^\text{boc}|$ when two BOCs are formed, and a lossless oscillation with frequencies $\mathcal{E}$ and $2\mathcal{E}$ when two BOCs and one BIC are formed. This result gives us a global map to control the excited-state population of the giant atom via engineering the separation of the two coupling points to the waveguide.

\begin{figure*}[tbp]
\includegraphics[width=2\columnwidth]{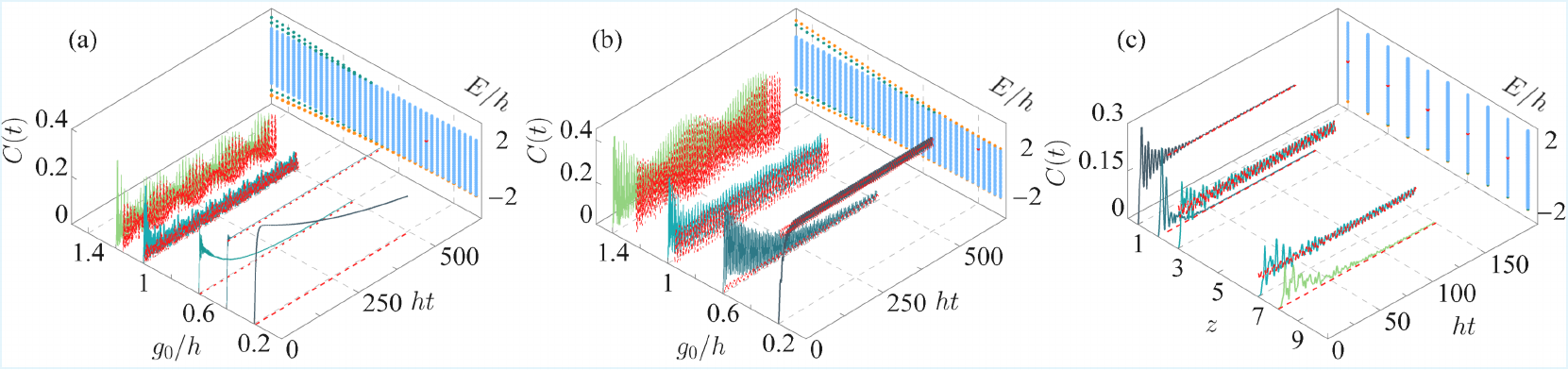}
\caption{Energy spectra and evolutions of the concurrence $C(t)$ in different (a) $g_0$ when $\Delta=0.16h$, $d=3$, and $z=1$ and (b) when $\Delta=1.04h$, $d=2$, and $z=1$ and (c) at different $z$ when $\Delta=0.36h$, $g_0=0.6h$, and $d=3$. The type-I BOCs are denoted by orange dots. The type-II BOCs are denoted by green dots. The BICs are denoted by red dots. The analytical results are plotted as the red dashed line. }\label{fig4}
\end{figure*}

\section{Two Giant Atoms}\label{sec_IV}
Next, we study the case of two giant atoms interacting with the waveguide through two resonators, i.e., $N=2$ and $M=2$, see Fig. \ref{fig1}. The two giant atoms are separated by $z$ resonators. The eigen equations derived from Eq. (\ref{eigene}) are
\begin{eqnarray}
&&Y_{e/o}(E) =E~\text{and}~ Y(E)\neq E,\label{solu1}\\
\text{or}~&&\int_{-2h}^{2h}d\omega{J_1(\omega)\over E-\omega}=0~\text{and}~ Y(E)=E,\label{solu2}
\end{eqnarray}
where $Y_{e/o}(E)\equiv \Delta+\int^{2h}_{-2h}d\omega{J_{0}(\omega)\pm J_{1}(\omega)\over E-\omega}$ and
\begin{equation}
J_{1}(\omega)={g_0^2\over 2\pi} {T_{z}(\tfrac{-\omega}{2h})+2T_{z+d}(\tfrac{-\omega}{2h})+T_{z+2d}(\tfrac{-\omega}{2h})\over \sqrt{4h^2-\omega^2}}.
\end{equation}
The limits of $Y_{e/o}(E)$ when $E\to \pm2h$ are given in Table \ref{lmt}. According to the criteria for forming the BOC, i.e., $Y_{e/o}(2h)>2h$ and $Y_{e/o}(-2h)<-2h$, the $Y_e(E)$ equation always has a type-I BOC below the band for any system parameters and the $Y_o(E)$ equation yields a type-II BOC below the band when
\begin{equation}
\Delta-(2z+d)g_0^2/h<-2h.\label{tbc1}
\end{equation}
For odd $d$, each of the $Y_{e/o}(E)$ equations gives a type-II BOC above the band under the same condition as Eq. (\ref{bsII}). For even $d$, the equation with the same parity as $z$ perpetually provides a type-I BOC above the band, while the equation with different parity from $z$ delivers a type-II BOC under the condition
\begin{equation}
\Delta+g_0^2(2z+d)/h>2h.\label{tbc2}
\end{equation}

\begin{table}
\begin{center} \caption{The limits of $Y_{e/o}(E)$ as $E$ approaches $\pm2h$ \cite{jia2024atom}. } \label{lmt}
\begin{tabular*}{\columnwidth}{@{\extracolsep{\fill}}c  c  c  c}
\toprule
      &$d\in\mathbb{O}$ &$d\in\mathbb{E}$ \& $z\in\mathbb{O}$ &$d\in\mathbb{E}$ \& $z\in\mathbb{E}$\\ \hline
  $Y_e(-2h)-\Delta$& \multicolumn{3}{c}{$-\infty$}\\
  $Y_o(-2h)-\Delta$& \multicolumn{3}{c}{$-(2z+d)g_0^2/h$}\\
  $Y_o(2h)-\Delta$ &  $dg_0^2/h$ & $\infty$& $(2z+d)g_0^2/h$\\
  $Y_e(2h)-\Delta$ &  $dg_0^2/h$ & $(2z+d)g_0^2/h$ & $\infty$\\
  \bottomrule
\end{tabular*}\end{center}\end{table}

The removable singularities of Eq. \eqref{solu1} determined by $J_0(\omega)\pm J_1(\omega)=0$ produce the BIC energies $E^{\text{bic}}_{e/o}=-2h\cos[l\pi/(d+z)]$, where the odd $l$ is from $Y_{e}(E)$ and even $l$ is from $Y_{o}(E)$. Those of Eq. \eqref{solu2} determined by $J_0(\omega)=0$ give $E^\text{bic}=-2h\cos[(2l+1)\pi/d]$. Substituting the former into Eq. \eqref{solu1} to analytically evaluate the integration, we obtain
\begin{equation}
    E^\text{bic}_{e/o}=\Delta+{2g_0^2\sin{l\pi d\over d+z}\over \sqrt{4h^2-E^{\text{bic}\,2}_{e/o}}}, \label{bictga}
\end{equation}which gives the condition for forming the BIC for the former case.  Substituting the latter into Eq. \eqref{solu2} to evaluate the integration, we obtain
\begin{equation}
    E^\text{bic}=\Delta. \label{mdds}
\end{equation}
However, satisfying the same form as the single-atom case, Eq. \eqref{mdds} results in two degenerate BICs. They are product states of the ground state of each giant atom and the single-atom BIC formed by the other giant atom and the environment.

The Laplace transform of Eq. \eqref{intfgr} for $N=2$ reads $\tilde{c}_{1/2}(s)=[\tilde{c}_e(s)\pm\tilde{c}_o(s)]/2$, where $\tilde{c}_{e/o}(s)=[s+i\Delta+\int_{-2h}^{2h}d\omega {J_0(\omega)\pm J_1(\omega)\over s+i\omega}]^{-1}$.
With the eigenenergies of the bound states obtained from Eq. \eqref{solu1} in hands, the long-time excited-state probability amplitudes are given by
\begin{eqnarray}
c_{1/2}(\infty)&=&\sum_{\alpha=\text{bic,boc}}[\sum_{j}{Z^\alpha_{e,j}\over2}e^{-iE_{e,j}^{\alpha}t}\nonumber\\&&\pm\sum_{j}{Z^\alpha_{o,j}\over 2}e^{-iE_{o,j}^{\alpha}t}],~~~~~\label{stdsl}
\end{eqnarray}
where $Z^\alpha_{e/o,j}=[\partial_s\tilde{c}_{e/o}(s)^{-1}|_{s\rightarrow -iE_{e/o,j}^\alpha}]^{-1}$ is the residue contributed by the $j$th bound-state energy $E^{\alpha}_{e/o,j}$ obtained from Eq. \eqref{solu1}. On the other hand, when the BIC with eigenenergy \eqref{mdds} is present, its contribution to the steady state of $c_1(t)$ should be estimated via Eq. (\ref{stdfm}). To reveal the effects of the bound states on the non-Markovian dynamics of the two giant atoms, we investigate the evolution of their entanglement. It is quantified by the concurrence $C(t)=\text{max}(0,\sqrt{\lambda_1}-\sqrt{\lambda_2}-\sqrt{\lambda_3}-\sqrt{\lambda_4})$, where $\lambda_i$ are the eigenvalues of $\rho(t)\tilde{\rho}(t)$ in decreasing order and $\tilde{\rho}(t)=(\tau_y\otimes\tau_y)\rho(t)^{\ast}(\tau_y\otimes\tau_y)$, with $\tau_y$ being the Pauli matrix \cite{PhysRevLett.80.2245}.

\begin{figure}[tbp]
\includegraphics[width=\columnwidth]{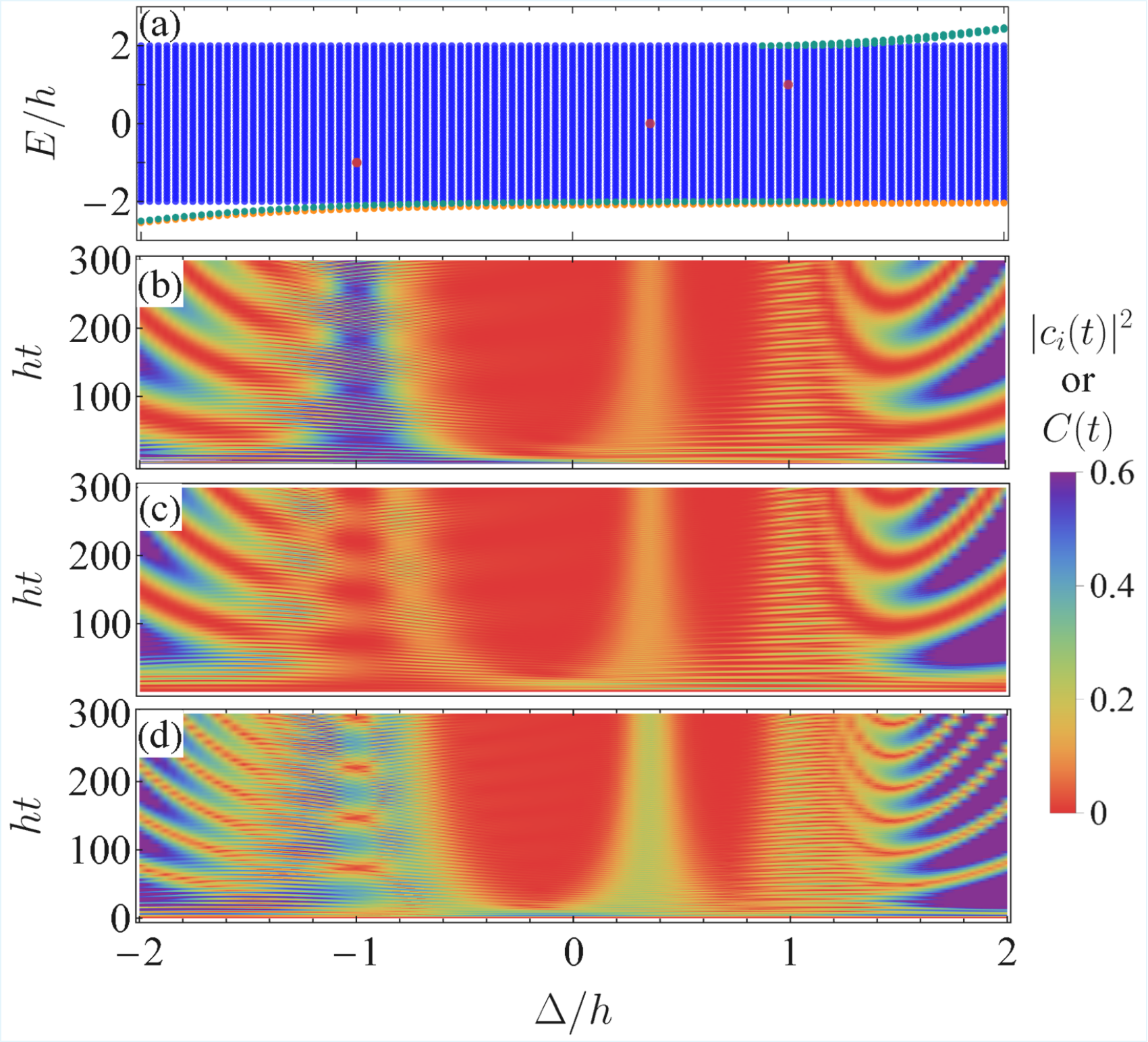}
\caption{(a) Energy spectrum and evolutions of (b) $|c_1(t)|^2$, (c) $|c_2(t)|^2$ and (d) $C(t)$ in different $\Delta$ when $g_0=0.6h$, $d=3$, and $z=3$. The type-I BOCs are denoted by orange dots. The type-II BOCs are denoted by green dots. The BICs are denoted by red dots.}\label{fig5}
\end{figure}

Figures \ref{fig4}(a) and \ref{fig4}(b) show the energy spectra as a function of the coupling strength $g_0$ when $d=3$ and $2$, respectively. For $d=3$, a type-I BOC contributed by the $Y_e$ equation always exists below the energy band. A type-II BOC below the band from $Y_o$ equation is present when $g_0\gtrsim0.66h$, which obeys Eq. \eqref{tbc1}. Two type-II BOCs from $Y_{e/o}$ emerge simultaneously above the energy band when $g_0 \gtrsim 0.87h$, which confirms the analytical criteria of Eq. \eqref{bsII}. For $d=2$, two type-I BOCs contributed by $Y_o$ and $Y_e$ are present above and below the band, respectively. Two type-II BOCs from the $Y_e$ equation when $g_0\gtrsim0.49h$, which obeys Eq. \eqref{tbc2}, and from the $Y_o$ equation when $g_0\gtrsim0.87h$, which obeys Eq. \eqref{tbc1}, are present. With increasing of the number of BOCs, the long-time concurrence jumps from the stability to a finite value to the multifrequency oscillations. In both cases, a BIC contributed by the $o$ component of Eq. \eqref{bictga} appears at $g_0=0.4h$ in \ref{fig4}(a) and $g_0=0.2h$ \ref{fig4}(b). It leads to a significant enhancement of the concurrence. The energy spectrum at different $z$ for $d=3$ is shown in Fig. \ref{fig4}(c). According to our analytical criteria, no BOC appears above the band regardless of the value of $z$ in this case. A type-I BOC is always present below the band. When $z \geq 3$, a type-II BOC obeying Eq. \eqref{tbc1} emerges below the energy band. Matching exactly to the values from Eq. \eqref{stdsl} in the long-time limit, the evolution of the concurrence clearly verifies the dominant role played by different numbers of bound states in the steady state.

Figure \ref{fig5} shows the energy spectrum and the evolution of $|c_{1/2}(t)|^2$ as a function of $\Delta$ when $d=z=3$. Being similar to Fig. \ref{fig4}(a), the type-I BOC always exists below the band. A type-II BOC below the band emerges when $\Delta<1.24h$, which obeys Eq. \eqref{tbc1}. Two nearly degenerate type-II BOCs are present above the band when $\Delta>0.92h$, which obeys Eq. \eqref{tbc2}. When $|\Delta|\le h$, the energies of the BOCs are very close to the band edge such that their contributed residues are too small to be observed in the dynamics. The exceptions occur at $\Delta = 0.36h$ and $h$, where the BICs obeying the $e$ and $o$ components of Eq. \eqref{bictga} are present and thus $|c_{1/2}(t)|^2$ approaches finite values with tiny-amplitude oscillation. When $|\Delta| >h$, the amplitude of the steady-state oscillation of $|c_{1/2}(t)|^2$ and $C(t)$ contributed by the BOCs increases, accompanying their energies moving away from the band edge. Another interesting result occurs at $\Delta=-h$, where the two degenerate BICs obeying Eq. \eqref{mdds} are present. They are
\begin{equation}
    |\text{BIC}_1\rangle=|g_1\rangle\otimes |\Phi_2^\text{bic}\rangle,~ ~|\text{BIC}_2\rangle=|\Phi_1^\text{bic}\rangle\otimes|g_2\rangle,
\end{equation}
where $|\Phi_j^\text{bic}\rangle$ is the single-atom BIC formed by the $j$th giant atom and the environment. Expanding the initial state $|\Psi(0)\rangle=|e_1,g_2,\{0_k\}\rangle$ in the complete basis of the total system, i.e., $|\text{BIC}_{1/2}\rangle$, $|\text{BOC}_{1/2}\rangle$, and the band states $|\Phi_\alpha\rangle$, we have
\begin{equation}
|\Psi(0)\rangle=x_2|\text{BIC}_2\rangle+\sum_{j=1}^2y_j|\text{BOC}_j\rangle+\sum_{\alpha\in\text{band}}z_\alpha |\Phi_\alpha\rangle, \label{ddvcmt}
\end{equation}where $x_j=\langle\text{BIC}_j|\Psi(0)\rangle$, $y_j=\langle\text{BOC}_j|\Psi(0)\rangle$, $z_\alpha=\langle\Phi_\alpha|\Psi(0)\rangle$. $|\text{BIC}_1\rangle$ does not contribute due to $x_1=0$. Then the dynamical evolution solution reads
\begin{eqnarray}
    |\Psi(t)\rangle&=&x_2e^{-i\Delta t}|\text{BIC}_2\rangle+\sum_{j=1}^2y_je^{-iE^\text{boc}_jt}|\text{BOC}_j\rangle\nonumber\\
    &&+\sum_{\alpha\in\text{band}}z_\alpha e^{-iE_\alpha t} |\Phi_\alpha\rangle.
\end{eqnarray}
In the long-time limit, the last term vanishes due to the out-of-phase interference. Thus, we have
\begin{eqnarray}
   \lim_{t\rightarrow\infty} c_1(t)&=&|x_2|^2e^{-i\Delta t}+\sum_{j=1}^2|y_j|^2e^{-iE_j^\text{boc}t},\\
   \lim_{t\rightarrow\infty} c_2(t)&=&\langle g_1,e_2,\{0_k\}|\sum_{j=1}^2y_je^{-iE^\text{boc}_jt}|\text{BOC}_j\rangle,
\end{eqnarray}
where $|\text{BIC}_2\rangle$ does not contribute to $c_2(t)$ due to $\langle g_1,e_2,\{0_k\}|\text{BIC}_2\rangle=0$. It indicates that $|c_1(\infty)|^2$ exhibits a tri-frequency oscillation, with frequencies proportional to $|\Delta-E_j^\text{boc}|$ and $|E_1^\text{boc}-E_2^\text{boc}|$, and $|c_2(\infty)|^2$ only exhibits a single-frequency oscillation, with a frequency proportional to $|E_1^\text{boc}-E_2^\text{boc}|$. Using the ansatz of Ref. \cite{PhysRevLett.122.073601}, we have the explicit form of the single-atom BIC with the eigenvalue $E^\text{bic}_l=-2h\cos[(2l+1)d/\pi]$ as
\begin{equation}
    |\Phi_1^\text{bic}\rangle =[\sin\theta \sqrt{2\over d}\sum_{n=0}^d \sin{(n\phi_l)}\hat{a}^\dagger_n+\cos\theta\hat{\sigma}^\dagger_1]|g_1,\{0_k\}\rangle,\label{dfdfc}
\end{equation}
with $\phi_l={(2l+1)\pi\over d}$, $\tan\theta={-g_0\over h\sqrt{2\over d}\sin\phi_l}$, and $n=0,...,d$ labeling the resonator from left to right inside the two coupling points. From Eq. \eqref{dfdfc}, we obtain
\begin{align}
x_2&=[1+\frac{dg_0^2}{{2h^2}\sin^2\phi_l}]^{-1/2}.
\end{align}
The parameter values used in Fig. \ref{fig5} and $\Delta=-h$ result in $x_2\approx 0.76$, which is much larger than $y_j$ according to the normalization of Eq. \eqref{ddvcmt}. Therefore, the amplitude of $|c_1(\infty)|^2$ containing the contribution of the BIC is much larger than the one of $|c_2(\infty)|^2$ without the contribution of the BIC when $\Delta=-h$. Although the BIC itself is a product state of the two giant atoms, its suppression of decoherence, together with the contribution from the BOCs, results in a higher long-time entanglement between the two giant atoms. All the above results demonstrate the one-to-one correspondence between the diverse steady-state behaviors of the giant atoms in the non-Markovian dynamics and the energy-spectrum feature of the composite giant-atom-environment system.

\section{Discussion and conclusion}\label{sec_V}
Our study focuses on giant atoms coupled to an LC-circuit resonators waveguide, but these findings can also be extended to systems composed of giant atoms interacting with rectangular waveguides \cite{Wang2022}, acoustic waveguides \cite{Manenti2017,science.1257219}, or transmission-line waveguides \cite{Kannan2020,PhysRevA.103.023710}. Although our work only considers one or two giant atoms interacting at two different sites, our conclusion is applicable to multiple atoms interacting with multiple points. In fact, the general formulation we developed in Sec. \ref{sec_II} is for $N$ giant atoms interacting with $M$ different sites of the one-dimensional coupled-resonator array. While the calculation may become more complicated for more atoms interacting with more points, and the quantitative condition for the formation of BOCs and BICs may differ, our conclusion that BOCs and BICs suppress the non-Markovian decoherence of giant atoms remains unchanged. Under current experimental conditions, the coupling strength $g_0/(2\pi)$ and the nearest-neighbor coupling strength $h/(2\pi)$ can reach the values of from $10$ to $100$ MHz, while the resonator frequency $\omega_c/(2\pi)$ and the giant-atom frequency $\Omega/(2\pi)$ typically lie in the range of $1$ to $10$ GHz \cite{PhysRevX.11.041043,PhysRevX.11.011015}. This large frequency separation strongly justifies the validity of the rotating-wave approximation. Furthermore, the coupling efficiency of the superconducting atom to the one-dimensional waveguide mode reaches $\beta \approx 0.999$ \cite{RevModPhys.95.015002}, implying that dissipation into other channels can be neglected in the analysis. Additionally, the existence of bound states and their induced dynamical effects has been confirmed experimentally in circuit QED systems \cite{Liu2017,PhysRevX.12.031036,Kannan2020}. This implies that our findings can be fully verified with existing experimental technologies.
Finally, our exact investigation in the full-parameter regime, which reveals the decoherence-suppression effects of both the BOCs and BICs on the giant atoms, is different from that of giant atoms interacting with a two-dimensional coupled cavity lattice in Ref. \cite{PhysRevResearch.6.043222}, which mainly focused on the impact of BICs on giant atom dynamics.

In summary, we have investigated the strongly coupled quantum electrodynamics between one and two giant atoms and a waveguide formed by an array of coupled-resonator waveguide. It has been revealed that the non-Markovian dynamics of the giant atoms exhibits rich behaviors. It is interesting to find that these rich behaviors are essentially determined by the feature of the energy spectrum of the composite system consisting of the giant atoms and the electromagnetic environment in the waveguide. When the energy spectrum only has a continuum band, all the giant atoms experience a complete decoherence. When a bound state either in or out of the continuum band is present in the energy spectrum, the excited-state probabilities of the giant atoms are stabilized to finite values. When more bound states in and out of the band are formed, each of the atoms tends to a lossless oscillation with frequencies proportional to the differences between their eigenenergies. The separations between the giant atoms and between the coupling points of each giant atom provide a useful dimension in engineering different types and numbers of bound states and thus controlling the decoherence dynamics of the giant atoms. Going beyond the widely used Born-Markov and WW approximations, our result refreshes our understanding of the strongly coupled giant-atom waveguide quantum electrodynamics and supplies us with an insightful mechanism to apply giant atoms in quantum interconnect.

\begin{acknowledgments}
The work is supported by the National Natural Science Foundation of China (Grants No. 12275109, No. 92576202, and No. 12247101), the Quantum Science and Technology-National Science and Technology Major Project (Grant No. 2023ZD0300904), the Natural Science Foundation of Gansu Province (Grants No. 26RCKA011 and No. 25JRRA799), the Fundamental Research Funds for the Central Universities (Grant No. lzujbky-2025-jdzx07), and the ‘111 Center’ (Grant No. B20063).
\end{acknowledgments}
\bibliography{ref.bib}
\end{document}